\documentclass[floats,twocolumn,prd,preprintnumbers,shortbibliography,10pt]{revtex4-1}
\usepackage[colorlinks=true,urlcolor=blue,linkcolor=blue,citecolor=blue]{hyperref}
\usepackage{slashed,color,amsmath,amssymb,mathrsfs}
\usepackage{amsfonts,epsfig,amssymb}
\usepackage{graphicx}
\usepackage{hyperref}
\usepackage{longtable}
\usepackage{array}
\usepackage{accents}
\usepackage{tensor}
\usepackage{booktabs}
\usepackage{multirow}
\usepackage{bm}
\usepackage{subfigure}

\newcommand{\sectitle}[1]{\vspace{.4cm}{\em #1.--}}
\allowdisplaybreaks

\begin{document}

\title[]{Nature of the $X(6900)$ in partial wave decomposition of $J/\psi J/\psi$ scattering }
\author{Qi Zhou$^{1,2}$}
\author{Di Guo$^{1,2}$}
\author{Shi-Qing Kuang$^{1,2}$}
\author{Qin-He Yang$^{1,2}$}
\author{Ling-Yun  Dai$^{1,2}$}\email{dailingyun@hnu.edu.cn}
\affiliation{$^{1}$ School of Physics and Electronics, Hunan University, Changsha 410082, China}
\affiliation{$^{2}$ Hunan Provincial Key Laboratory of High-Energy Scale Physics and Applications, Hunan University, Changsha 410082, China}

\date{\today}

\begin{abstract}
In this letter, we perform partial wave decomposition on coupled-channel scattering amplitudes, $J/\psi J/\psi$-$J/\psi \psi(2S)$-$J/\psi \psi(3770)$, to study the resonance appears in these processes. Effective Lagrangians are used to describe the interactions of four charmed vector mesons, and the scattering amplitudes are calculated up to the next-to-leading order. Partial wave projections are performed, and unitarization is implemented by Pad\'e approximation. Then we fit the amplitudes to the $J/\psi J/\psi$ invariant mass spectra measured by LHCb and determine the unknown couplings. The pole parameters of the $X(6900)$ are extracted as $M=6861.0^{+6.3}_{-8.8}$~MeV and $\Gamma=129.0^{+5.6}_{-3.4}$~MeV. Our analysis implies that its quantum number prefers to be $0^{++}$. The pole counting rule and phase shifts show that it is a normal Breit-Wigner resonance and,  hence, should be a compact tetraquark. 
\end{abstract}

\maketitle

\sectitle{Introduction}
\label{sec:Intro}
Searching for multiquark states plays an important role in understanding QCD. Once its existence is confirmed, the inner structure of hadrons would be changed: They can be composed of not only the traditional components,  $\bar{q}q$ for meson and $qqq$ for baryon as suggested by the quark model \cite{Gell-Mann:1964ewy,Zweig:1964ruk,Zweig:1964jf}, but also $\bar{q}\bar{q}qq$ and $\bar{q}qqqq$, etc. For some recent reviews on this topic, we refer to Refs.~\cite{Guo:2017jvc,Brambilla:2019esw,Yao:2020bxx,Chen:2022asf}.  
In the past decade, some hidden-charm resonances were discovered in the spectra of $J/\psi \pi$ and/or $J/\psi p$, e.g. $Z_c$ states by BESIII \cite{Ablikim:2013mio} and Belle \cite{Liu:2013dau} and $P_c$ states by LHCb \cite{Aaij:2015tga,Aaij:2019vzc}. These may start a new era of particle physics as the resonances mentioned above contain at least four or five quark components.
Very recently, big progress in this field has been made by LHCb again, where a fully heavy tetraquark candidate is found~\cite{LHCb:2020bwg}. This narrow structure near 6900~MeV/$c^2$, labeled as $X(6900)$, was found in $J/\psi J/\psi$ invariant mass spectra with statistical significance of the signal more than 5$\sigma$. The mass and width are determined to be either
\begin{eqnarray}
m[X(6900)]&=&6905\pm 11\pm7 MeV/{c^2}\,, \nonumber\\
\Gamma[X(6900)]&=&80\pm 19\pm33 MeV/{c^2}\,, \nonumber
\end{eqnarray}
or
\begin{eqnarray} 
m[X(6900)]&=&6886\pm 11\pm11 MeV/{c^2}\,, \nonumber\\
\Gamma[X(6900)]&=&168\pm 33\pm69 MeV/{c^2}\,, \nonumber
\end{eqnarray}  
with different treatments on the contribution of nonresonant single-parton scattering continuum \cite{LHCb:2020bwg}. 
This fueled further interests of the community, see e.g., Refs.~\cite{Dong:2020nwy,Wang:2020wrp,Gong:2020bmg,Cao:2020gul,Guo:2020pvt,Liang:2021fzr,Wang:2020gmd,Ke:2021iyh,Chen:2020xwe,Wang:2022jmb}. 
Several natural following questions would be: What is the quantum number of this state, and What is the structure? These are the critical concerns of our paper. 

One needs partial wave decomposition to extract the information and the quantum number of the resonance \cite{Dai:2014lza, Dai:2014zta}. Furthermore, phase shifts help study hadronic scattering as well as resonances appearing in the intermediate states\footnote{For instance, in the $\pi\pi, \pi K$ scatterings, the phase shifts~\cite{Hyams:1973zf, Aston:1987ir} help to confirm the existence of the light scalars, $\sigma,\kappa$~\cite{Ishida:1995xx,Ishida:1997wn,Xiao:2000kx,Zhou:2004ms}.}. In another aspect, the pole counting rule \cite{Morgan:1992ge,Dai:2011bs} helps to distinguish the inner structure of resonances: molecule or Breit-Wigner-type resonance. Combining these methods, we can comprehensively analyze the property of the $X(6900)$.

\sectitle{Formalism} \label{Sec:II}
To study the $X(6900)$, we focus on the energy region from $2M_{J/\psi}$ to 7200~MeV.
In our analysis, we consider the triple-channel scatterings,  $J/\psi J/\psi$-$J/\psi \psi(2S)$-$J/\psi \psi(3770)$, as the thresholds of $J/\psi \psi(2S)$ and $J/\psi \psi(3770)$ are the closest to the resonant structure around 6900~MeV. As a comparison, we also consider coupled-channel scatterings,  $J/\psi J/\psi$-$J/\psi \psi(2S)$.  The interactions of four heavier vectors, e.g. $\psi(2S)\psi(2S)\psi(3770)\psi(3770)$, are ignored as the thresholds of $\psi(2S)\psi(2S)$ and $\psi(3770)\psi(3770)$ are much heavier. The $\eta_c\eta_c$ and $h_ch_c$ channels are suppressed by heavy quark spin symmetry (HQSS) and are neglected~\cite{Dong:2020nwy,Gong:2022hgd}, too.  
The effective Lagrangians of interactions are 
\begin{eqnarray}
    \mathcal{L}&=&c_1V_{\mu} V_{\alpha} V^\mu V^\alpha+c_2 V_{\mu} V_{\alpha} V^\mu V^{\prime\alpha}+c_3V_{\mu} V^\prime_{\alpha} V^\mu V^{\prime\alpha} \nonumber\\
    &+&
    c_4V_{\mu} V^{\prime\mu} V_{\alpha} V^{\prime\alpha}+c_5V_{\mu} V_{\alpha} V^\mu V^{\prime\prime\alpha} +c_6V_{\mu} V^{\prime\prime}_{\alpha} V^\mu V^{\prime\prime\alpha}\nonumber\\
    &+&c_7V_{\mu} V^{\prime\prime\mu} V_{\alpha} V^{\prime\prime\alpha}+c_8V_{\mu} V^\prime_{\alpha} V^\mu V^{\prime\prime\alpha}+c_9V_{\mu} V^{\prime\mu} V_{\alpha} V^{\prime\prime\alpha}\,,\nonumber\\ \label{eq:lagran}
\end{eqnarray}
where $V$, $V^\prime$, and $V^{\prime\prime}$ represent for $J/\psi$, $\psi(2S)$, and $\psi(3770)$, respectively.  It satisfies the discrete symmetries, C, P, and T. 
These effective Lagrangians are indeed the same as the leading-order (LO) Lagrangians constructed from HQSS \cite{Casalbuoni:1996pg}. For example, one has $\mathcal{L}_{\rm HQSS}^{LO}= g_1 \langle J\bar{J}J\bar{J}\rangle =2N_C g_1 V_{\mu} V_{\alpha} V^\mu V^\alpha$. 
The higher-order Lagrangians will be suppressed by $1/m_Q$ \footnote{Higher-order Lagrangians with derivatives will be suppressed by HQSS, too. 
The momentum coming from derivatives will be carried mainly by the velocity $v=(1,\vec{0})$, and one has $v\cdot V^{(','')}=0$.}. 

\begin{figure}[h]
   \centering
   \begin{center}
   \vspace{-0.cm}
   {\includegraphics[width=0.45\textwidth,height=0.16\textheight]{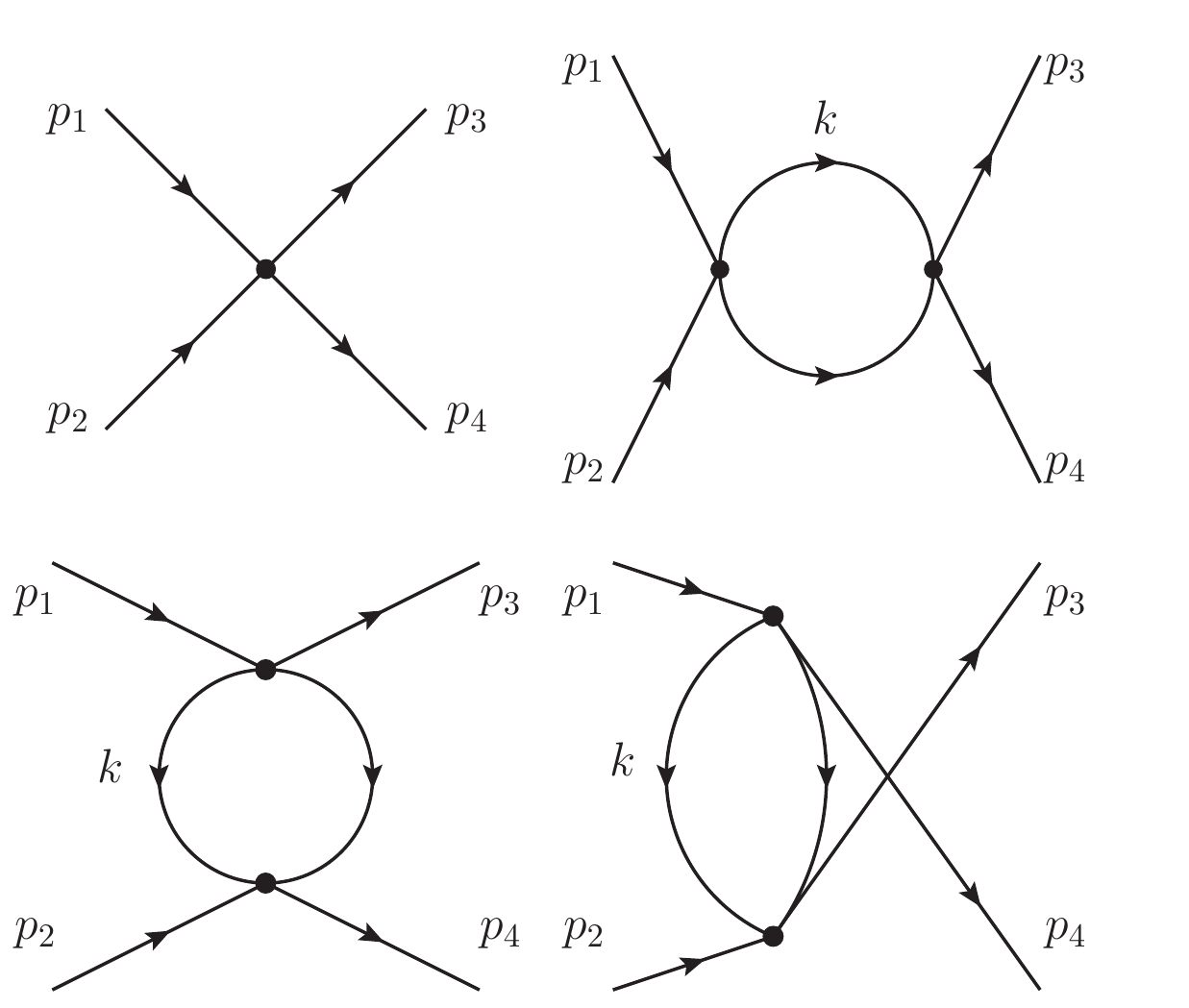}} 
\end{center}
   \caption{Feynman diagrams of the scattering amplitudes. The intermediate states include $J/\psi J/\psi$, $J/\psi \psi(2S)$, and $J/\psi \psi(3770)$.   
   \label{Fig:Feynman diagram} }
\end{figure}

With Eq.~(\ref{eq:lagran}) we calculate the scattering amplitudes up to next-to-leading order (NLO)\footnote{Notice that the NLO results would supply not only higher-order energy-dependent potentials, but also the left-hand cut contributions. }, see Fig.\ref{Fig:Feynman diagram}.
The scattering amplitudes $T^{ij}$ can be expressed as
\begin{eqnarray}
    T^{ij}&=&F^{ij}_{(a)}(\varepsilon_1\cdot\varepsilon_2)(\varepsilon_3^*\cdot\varepsilon_4^*)+F^{ij}_{(b)}(\varepsilon_1\cdot\varepsilon_3^*)(\varepsilon_2\cdot\varepsilon_4^*)\nonumber\\
    &+&F^{ij}_{(c)}(\varepsilon_1\cdot\varepsilon_4^*)(\varepsilon_2\cdot\varepsilon_3^*) \,, \label{eq:amp}
\end{eqnarray}
where the superscripts $i$ and $j$ are channel labels, with the numbers 1, 2, and 3 specified as $J/\psi J/\psi$, $J/\psi \psi(2S)$ and $J/\psi \psi(3770)$, respectively. The subscripts 1, 2, 3, and 4 of the polarization vectors are labels of mesons. The subscripts  (a, b, and c) are used to tag form factors related to different polarization structures.

To clarify the quantum number of the $X(6900)$, partial wave projections are needed. 
The partial wave amplitudes can be obtained via the decomposition of helicity amplitudes \cite{Martin1970}:
\begin{eqnarray}\label{eq:pwh}
   T_{\mu_1\mu_2;\mu_3\mu_4}^{J,ij}(s)=\frac{1}{ 32\pi N}\int_{-1}^1  T_{\mu_1\mu_2;\mu_3\mu_4}^{ij}(s,z_s)d_{\mu\mu^\prime}^J(\theta_s)dz_s \,. \nonumber\\
\end{eqnarray}
where $s$ is the Mandelstam variable, $s=(p_1+p_2)^2$. $\theta_s$ is the scattering angle in the center of mass frame (c.m.f.) in the $s$ channel, and $z_s=\cos\theta_s$. $\mu=\mu_1-\mu_2$ and $\mu^\prime=\mu_3-\mu_4$ are the initial and final states' helicity, respectively.  $d_{\mu\mu\prime}^J(\theta_s)$ is the standard Wigner function for rotation. $N$ is the normalization factor caused by the property of identical particles, with $N=2$ for $T^{11}$ amplitude, $N=\sqrt{2}$ for $T^{12,13}$ amplitudes, and $N=1$ for others. 
For each angular momentum $J$, parity and time reversal conservation can reduce the number of independent helicity amplitudes. 
With parity conservation, one has 41 independent helicity amplitudes for $T_{\mu_1\mu_2;\mu_3\mu_4}^{J,ij}(s)$. With time reversal, the elastic scattering amplitudes will be reduced again, resulting in 25 independent ones.
Furthermore, one needs to transfer the amplitudes from $|JM\mu_1\mu_2\rangle$ representation into the $|JMLS\rangle$ one~\cite{Martin1970},
\begin{eqnarray}
&&T_{\mu_1\mu_2;\mu_3\mu_4}(s,z_s) = 16\pi N\sum_{J} (2J+1) d_{\mu\mu'}^J(\theta_s) \nonumber\\
&&\sum_{LS,L'S'}\frac{\sqrt{(2L+1)(2L'+1)}}{2J+1}\left\langle LS0\mu|J\mu\right\rangle \left\langle J\mu'| L'S'0\mu'\right\rangle \nonumber\\
 &&  \left\langle s_1s_2\mu_1,-\mu_2|S\mu\right\rangle \left\langle S'\mu' |s_3s_4\mu_3,-\mu_4 \right\rangle 
  T^J_{LS,L'S'}  \,. \label{eq:pwLS}
\end{eqnarray}
The Clebsch-Gordan coefficients can be found in PDG \cite{ParticleDataGroup:2020ssz}.

For $J/\psi J/\psi$ system, the eigenvalues of charge conjugation and parity transformations are given by $C=(-1)^{L+S}=+, P=(-1)^L$. Higher partial waves can be ignored, and only the lowest ones with $L=0,1$ are considered. Therefore, five partial waves are left: $S$-waves, $0^{++}$ and $2^{++}$; $P$-waves, $0^{-+}$, $1^{-+}$, and $2^{-+}$ \footnote{Higher-order Lagrangians with derivatives could contribute to the higher partial waves. However, they are suppressed by HQSS. Further, a reliable description of them relies on the angular distributions, and it is expected that future experiments can supply more measurements to ensure a more detailed analysis.}. 
See Table \ref{tab:JPC}. 
\begin{table}[htbp]
{\footnotesize
\begin{tabular}{ccccc}
\hline\hline
$L$  \rule[-0.2cm]{0cm}{6mm}    & $S=0$  &  $S=1$   &$S=2$   \\ \hline
$0$  \rule[-0.2cm]{0cm}{6mm}    & ${0}^{{++}}~(^1S_0)$ & $\cdots$   &  ${2}^{{++}}~(^5S_2)$ \\
$1$   \rule[-0.2cm]{0cm}{6mm}   & $\cdots$ &${{0}^{{-+}}}~(^3P_0)\quad{{1}^{{-+}}}~(^3P_1)\quad{{2}^{{-+}}}~(^3P_2)$   &  $\cdots$ \\
 \hline\hline
\end{tabular}
\caption{Quantum number $J^{PC}$ of $J/\psi J/\psi$ partial waves. The number in the bracket is in the form of $^{2S+1}L_J$. The $\cdots$ denotes other possible quantum numbers of $J/\psi \psi(2S)$ and $J/\psi \psi(3770)$ waves that are neglected as they are forbidden in the $J/\psi J/\psi$ system.  \label{tab:JPC}  }
}
\end{table}
Notice that there is no coupling between partial waves with different orbit momentum (for example, $^1S_0-^5D_0$). 
The partial wave amplitudes are given as 
\begin{eqnarray}
   T_{^1S_0}^{ij}(s)&=&\frac{2}{3}T_{++++}^{0,ij}(s)+\frac{2}{3}T_{++--}^{0,ij}(s)-\frac{2}{3}T_{++00}^{0,ij}(s)\nonumber\\
   &-&\frac{2}{3}T_{00++}^{0,ij}(s)+\frac{1}{3}T_{0000}^{0,ij}(s)\,, \nonumber\\
  T_{^1S_0}^{ii}(s)&=&\frac{2}{3}T_{++++}^{0,ii}(s)+\frac{2}{3}T_{++--}^{0,ii}(s)-\frac{4}{3}T_{++00}^{0,ii}(s) \nonumber\\
  &+&\frac{1}{3}T_{0000}^{0,ii}(s)\,, 
\label{eq:pw25}
\end{eqnarray}
where we give only the expressions of $^1S_0$ waves for simplicity. 
As can be checked, the partial wave scattering amplitudes obtained from Lagrangians of Eq.(\ref{eq:lagran} ) can produce the correct threshold behavior $T^J_{LS,L'S'}\varpropto p_{cm}^L {p'}_{cm}^{L'}$, with 
$p_{cm}$, $p'_{cm}$ the modulus of the three momenta for initial and final states, respectively, in c.m.f., e.g., $p_{cm}=\sqrt{\left(s-(m_{a}+m_{b})^2\right)\left(s-(m_{a}-m_{b})^2\right)}/2$.
For instance, one has  $T_{^3P_1}^{11,LO} (s)=-c_1 (s-4m_V^2)/(12\pi m_V^2)$ and $T_{^3P_1}^{12,LO} (s) =-c_2 (m_V+m_{V'}) p_{cm} p'_{cm}/(12\sqrt{2}\pi m_V^2 m_{V'} )$.

The unitarity  of the partial wave amplitudes in terms of the $|JMLS\rangle$ representation is given as  \cite{Martin1970,Chung:1971ri,Oller:2019opk}
\begin{eqnarray}
&&   \left\langle L'S'| T^J | LS\right\rangle-\left\langle L'S' |T^{J\dagger}|LS\right\rangle \nonumber\\
=&&i\frac{4|\vec{p}~^{\prime\prime}|}{E_{cm}''} \sum_{L''S''}\left\langle  L'S'|T^{J\dagger}|L''S''\right\rangle\left\langle  L''S''|T^J|LS\right\rangle\,, \label{eq:unit;LS}
\end{eqnarray}
where the quantum number $J$ and $M$ in the kets has been ignored for simplicity. 
The summation symbol in Eq.\eqref{eq:unit;LS} can be removed since $L\leq 1$. On the other hand, the coupled-channel scatterings of $J/\psi J/\psi$-$J/\psi \psi(2S)$-$J/\psi \psi(3770)$ are included, and finally, the unitarity relation is given as 
\begin{eqnarray}
  {\rm Im} T^{ij}_{JLS}&=&\sum_{k=1}^a T^{ik}_{JLS}~\rho_k ~T^{kj~*}_{JLS}\,, \label{eq:unit;3}
 \end{eqnarray}
where $\rho_k$ is the phase space factor for the $k$th channel $\rho_k(s)=2|\vec{p}_k|/\sqrt{s}$ \cite{Kuang:2020bnk}. In the summation symbol, $a = 2$ is for coupled-channels case, and $a = 3$ for triple-channel case. 
The scattering amplitudes given in Eq.~(\ref{eq:pw25}) are calculated in the spirit of perturbation theory and work only in the low-energy region. 
Pad\'e  approximation \cite{Truong:1988zp,Dai:2011bs,Dai:2012kf} is applied to extend the amplitudes to a higher-energy region concerning for unitarity, 
\begin{eqnarray}
   T=T^{LO}\cdot[T^{LO}-T^{NLO}]^{-1}\cdot T^{LO} \,, \label{eq:pade}
\end{eqnarray}
where it is written in matrix form. Equation~\eqref{eq:pade} can restore the perturbation amplitudes up to NLO in the low-energy region. Similar approaches, such as the inverse amplitude method, have been applied successfully in unitarizing chiral amplitudes \cite{Dobado:1996ps,Oller:1997ng,GomezNicola:2001as}.
Three partial wave scattering amplitudes, $^1S_0$, $^5S_2$, and $^3P_1$, are unitarized with this approach. 
For the partial wave amplitudes of $^3P_0$ ($0^{-+}$) and $^3P_2$ ($2^{-+}$), the tree diagrams vanished, and their loop corrections are small. Hence, we do not perform unitarization on them but use the perturbative amplitudes instead. 
With these five partial waves, one can extract the pole information and determine the quantum number of the resonance. See discussions below.

\sectitle{Fit results and discussion}\label{Sec:III}
We fit the partial wave amplitudes to the $J/\psi J/\psi$ invariant mass spectra, and the couplings of the effective Lagrangians can be fixed. 
To specify the contribution to the invariant mass spectra of each channel [$T^{i1}(s)$],
we simply assume that the $i$th channel contributes a ratio, $\alpha_i$, with the normalization $\sum_i\alpha_i^2\equiv 1$. One then has such a formula to fit the $J/\psi J/\psi$ invariant mass spectra \cite{Dai:2021wxi}
\begin{eqnarray}
   \frac{d~{\rm events}}{d\sqrt{s}}&=&\tilde{N}~ p_{cm}(s) \sum_{\footnotesize \mu_1\mu_2 \mu_3\mu_4}   \int_{-1}^{1}dz_s \nonumber\\ 
  && |\sum_{i=1}^{a}\alpha_i T^{i1}_{\footnotesize \mu_1\mu_2 \mu_3\mu_4}  (s,z_s)|^2 \,,        \label{eq:events}
\end{eqnarray} 
where the superscript  $i$ and 1 are channel labels. $\tilde{N}$ is a normalization factor, with other factors such as the integration on the azimuthal angle $\phi$ absorbed. 
Though $\tilde{N}$ is correlated with $\alpha_k(s)$, and they will be dependent on each other in the fitting procedure, this problem has been solved due to the normalization, $\sum_i\alpha_i^2\equiv 1$. 
With  Eq.~(\ref{eq:pw25}), the integration on the square of the helicity amplitudes can be expressed by partial wave amplitudes:
\begin{eqnarray}
  &&\sum_{\footnotesize \mu_1\mu_2 \mu_3\mu_4}     \int_{-1}^{1}|\sum_{i=1}^{a}\alpha_i T^{i1}_{\footnotesize \mu_1\mu_2 \mu_3\mu_4}  (s,z_s)|^2 dz_s \nonumber\\
  =&&512\pi^2
  \left[~|F^{1}_{^1S_0}(s)|^2 
   +5|F^{1}_{^5S_2}(s)|^2+|F^{1}_{^3P_0}(s)|^2 \right.\nonumber\\  &&\;\;\;\;\;\;\;\left.+3|F^{1}_{^3P_1}(s)|^2+5|F^{1}_{^3P_2}(s)|^2~
  \right] \, , 
  \label{eq:mod;T}
\end{eqnarray}
where $F^1_{JLS}(s)=\sum_{i=1}^a\alpha_i N_i T^{i1}_{JLS}(s) $, with $N_i$ given in Eq.(\ref{eq:pwh}). This amplitude is consistent with the Au-Morgan-Pennington method \cite{Au:1986vs,Dai:2014lza,Dai:2016ytz}, where contributions of the left-hand cut and distant right-hand cut are absorbed into $\alpha_i$ concerning for coupled-channel unitarity and final state interactions. 
At last, we consider both coupled-channels scattering, $J/\psi J/\psi$-$J/\psi \psi(2S)$ (Fit.I) and triple-channel scatterings, $J/\psi J/\psi$-$J/\psi \psi(2S)$-$J/\psi \psi(3770)$ (Fit.II). Indeed, it is found that the third channel $J/\psi \psi(3770)$ contributes only a bit. See discussions below.

The input parameters, such as the masses of the particles, are taken from PDG \cite{ParticleDataGroup:2020ssz}, which are given as: $m_{J/\psi}=3096.9$~MeV, $m_{\psi(2S)}=3686.1$~MeV, $m_{\psi(3770)}=3770.7$~MeV. The renormalization scale of one-loop amplitudes is taken as $\mu=1$~GeV. The other parameters, the couplings of the effective Lagrangians and the normalization factor, are fixed by fits, with the MINUIT procedure \cite{James:1975dr}. See Table \ref{tab:para}.
\begin{table}[htbp]
   {\footnotesize  
      \begin{tabular}{ccc}
      \hline\hline
      Parameter  & Fit.I  &  Fit.II                           \\ \hline 
      $c_1$      & $-0.1232_{-0.0001}^{+0.0001}$  & $-0.1263_{-0.0002}^{+0.0007}$  \\[0.75mm]
      $c_2$      & $-0.5359_{-0.0001}^{+0.0021}$   & $-0.5859_{-0.0001}^{+0.0001} $ \\[0.75mm]
      $c_3$      & $-0.3250_{-0.0001}^{+0.0171}$  & ~~$0.1607_{-0.0013}^{+0.0024} $  \\[0.75mm]
      $c_4$      & $-0.6277_{-0.0002}^{+0.0234}$  &$-1.0326_{-0.0022}^{+0.0055}   $\\[0.75mm]
      $c_5$      &  $\cdots$   & $ -0.0707_{-0.0001}^{+0.0001}$ \\[0.7mm]
      $c_6$      &  $\cdots$    & $-0.2808_{-0.0003}^{+0.0006}  $  \\[0.75mm]
      $c_7$      &  $\cdots$    & ~~$0.5998_{-0.0003}^{+0.0007} $ \\[0.75mm]
      $c_8$      &  $\cdots$    &~~$0.2361_{-0.0001}^{+0.0003}  $    \\[0.75mm]
      $c_9$      &  $\cdots$    &$-0.2162_{-0.0001}^{+0.0007}  $  \\[0.75mm]
      $\tilde{N}$ & ~~$1.2589_{-0.0850}^{+0.6284}$   &~~$3.2546_{-0.2999}^{+1.5452}$   \\[0.7mm]
      $\alpha_1$ & ~~$0.3691_{-0.0052}^{+0.0104}$  & ~~$0.3307_{-0.0254}^{+0.0529}$   \\[0.75mm]    
     $\alpha_2$ & $-0.9294_{-0.0022}^{+0.0089}$  & $-0.7711_{-0.0511}^{+0.1072}$   \\[0.75mm]
      $\alpha_3$ & $\cdots$  & $-0.5441_{-0.0633}^{+0.1085} $       \\[0.75mm]
      $\chi^2_{d.o.f.}$    &   1.29        &1.28      \\
      \hline\hline
      \end{tabular}
      \caption{Parameters of our solution. Unit of the normalization factor $\tilde{N}$ is $10^{-4}{\rm MeV}^{-2}$. The uncertainties of the parameters are taken from bootstrap.  \label{tab:para}  }
   }       
      \end{table}
The errors of the parameters are mainly from bootstrap \cite{Efron:1979bxm}, where they are counted by varying the experimental data within its uncertainty by multiplying a normal distribution function. The uncertainties from MINUIT are much smaller and thus ignored.

\begin{figure}[htbp]
   \centering
   \includegraphics[width=0.98\linewidth, height=0.25\textheight]{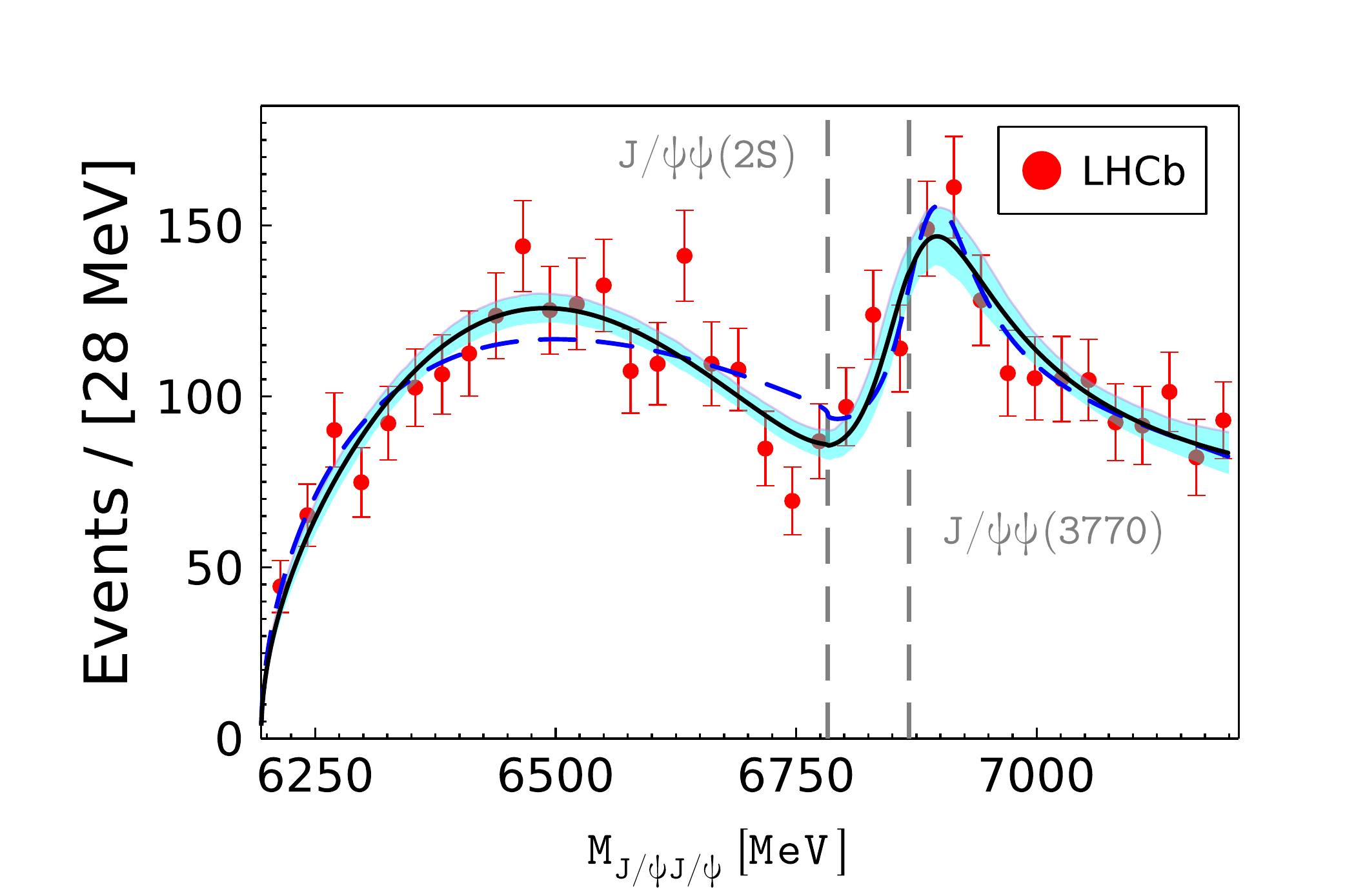}
   \caption{Fit to the invariant mass spectra of LHCb \cite{LHCb:2020bwg}. The dashed blue line is for Fit.~I and the solid black line is for Fit.~II. The cyan band is the uncertainty of Fit.~II estimated from the bootstrap method within 1$\sigma$. \label{Fig:fit}}
\end{figure}

As presented in Table \ref{tab:para}, the $\chi^2_{d.o.f.}$ of Fits.~I and II are similar to each other, while the solution of the triple-channel fits a bit better in the energy region from 6400 to 6800~MeV, see Fig.\ref{Fig:fit}. 
This is not surprising, as Fit.~II includes more contributions from different channels. 
Nevertheless, both solutions fit the data around the $X(6900)$ rather well. 

\begin{figure}[htbp]
\centering
\includegraphics[width=0.98\linewidth, height=0.25\textheight]{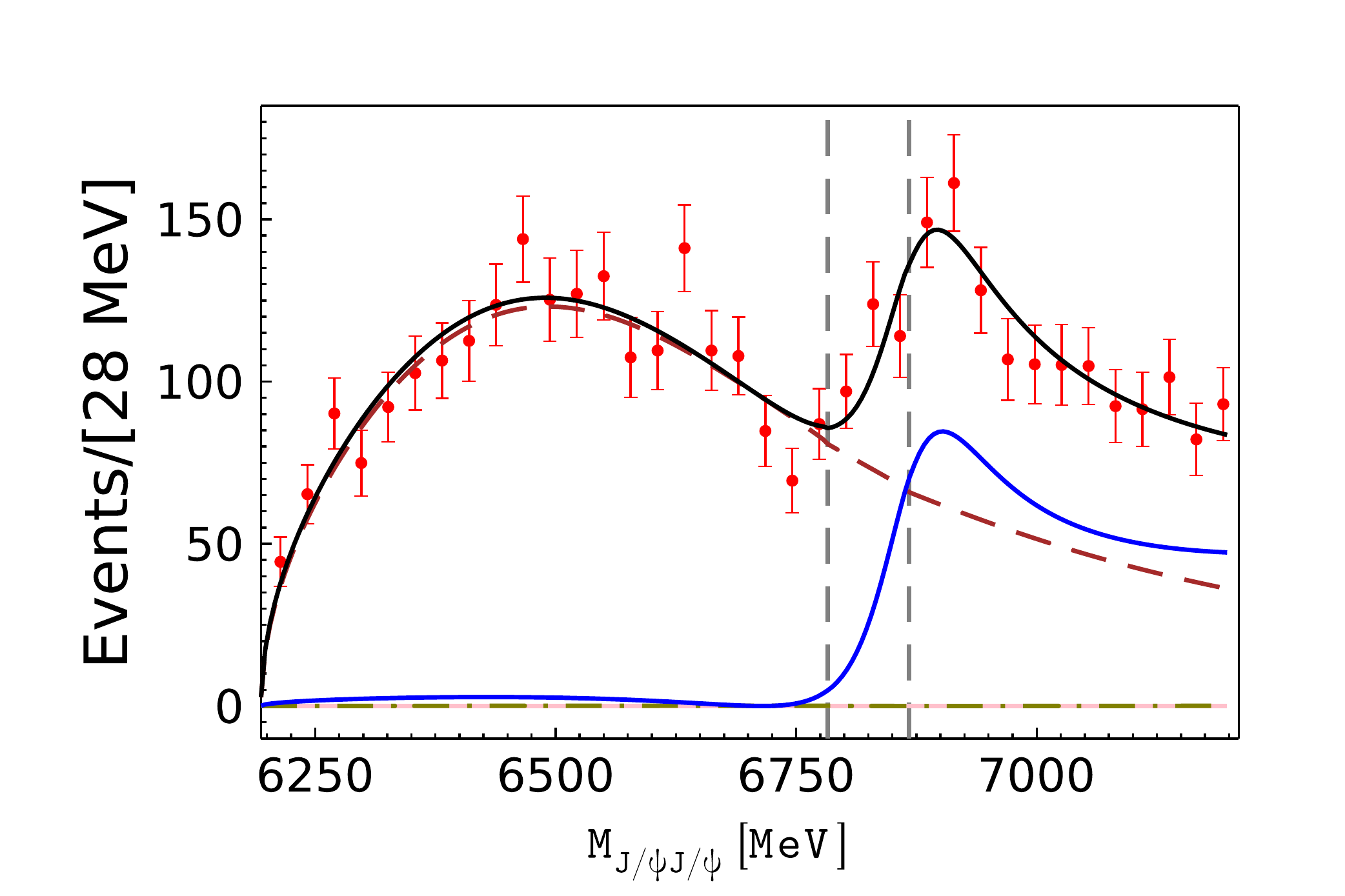}
\caption{Individual contribution of each partial wave for the triple-channel case. The solid blue, dashed brown, dash-dotted olive, dotted gray, and solid pink lines are for $^1S_0$, $^5S_2$,  $^3P_1$, $^3P_0$, and $^3P_2$, respectively. \label{Fig:ind} }
\end{figure}

The individual contribution of each partial wave of Fit.~II is shown in Fig.\ref{Fig:ind}, and that of Fit.~I is quite similar, and we do not plot it here.  
As can be seen, the $^1S_0$ wave contributes a resonant structure around 6900~MeV, and the $^5S_2$ wave contributes a smooth background in the whole energy region. This suggests that the $X(6900)$ is more likely to be $^1S_0$ ($0^{++}$) state. As expected, the $^3P_1$ wave contributes a little background. The other two waves, $^3P_0$ and $^3P_2$ are relatively small and can be ignored. 

To study the property of the $X(6900)$ more carefully, we extract out pole locations, i.e., the masses and widths of the resonance from the scattering amplitudes. First, the amplitudes are continued into the complex-$s$ plane. Then the poles are searched in each partial wave. The pole information for coupled and triple channels is shown in Table \ref{tab:poles;2} and Table \ref{tab:poles;3}, respectively.
\begin{table}[t]
   {\footnotesize
   \begin{tabular}{|c|c|c|c|}
   \hline
   \rule[-0.2cm]{0cm}{7mm}RS         &\rule[-0.2cm]{0cm}{7mm} Pole location (MeV)   &    $|g_1|$(MeV)&    $|g_2|$(MeV)\\
   \hline
   \rule[-0.2cm]{0cm}{7mm} II  (- +) & \rule[-0.2cm]{0cm}{7mm} $6886.8_{-4.6}^{+8.3}$-$i17.6_{-0.3}^{+2.5} $   &$998.6_{-4.2}^{+43.1}$      &$688.1_{-2.1}^{+6.9}$ \\
   \hline
   \rule[-0.2cm]{0cm}{7mm} III(- -)  & \rule[-0.2cm]{0cm}{7mm} $6884.2_{-4.0}^{+8.2}$-$i29.1_{-0.1}^{+1.5}  $  & $992.0_{-4.2}^{+20.0}$        &$680.9_{-1.9}^{+4.5}$  \\
   \hline
   \end{tabular}
   \caption{\label{tab:poles;2} Pole locations for Fit.I. } 
   }
   \end{table}
\begin{table}[t]
{\footnotesize
   \begin{tabular}{|c|c|c|c|c|}
   \hline
   \rule[-0.2cm]{0cm}{5mm}\multirow{2}{*}{RS}                                       & \multirow{2}{*}{\begin{tabular}[c]{@{}c@{}}Pole location \\ (MeV)\end{tabular}}     & \multirow{2}{*}{$|g_1|$(MeV)}  & \multirow{2}{*}{$|g_2|$(MeV)}  &  \multirow{2}{*}{$|g_3|$(MeV)}   \\
   &        &      &     &    \\ \hline
   \rule[-0.12cm]{0cm}{4.8mm} II & $6872.7_{-8.6}^{+6.0}$  & \multirow{2}{*}{$1352.7_{-11.7}^{+28.1}$}  & \multirow{2}{*}{ $946.5_{-8.0}^{+18.8}$}  & \multirow{2}{*}{$14.4_{-0.3}^{+0.9}$} \\
   \rule[-0.12cm]{0cm}{4.8mm} (- + +) &   -$i46.5_{-1.0}^{+2.1} $     &      &      &    \\ \hline
   \rule[-0.12cm]{0cm}{4.8mm} III & $6861.0_{-8.8}^{+6.3}$  & \multirow{2}{*}{$ 1326.2_{-11.3}^{+24.7}$} & \multirow{2}{*}{$917.9_{-7.9}^{+15.1}$} & \multirow{2}{*}{$16.0_{-0.4}^{+0.8}$} \\
   \rule[-0.12cm]{0cm}{4.8mm} (- - +) &    -$i64.5_{-1.7}^{+2.8} $     &           &       &    \\ \hline
   \rule[-0.12cm]{0cm}{4.8mm} IV & $6861.0_{-8.8}^{+6.3}$ & \multirow{2}{*}{$ 1322.9_{-14.6}^{+26.2}$} & \multirow{2}{*}{$915.6_{-9.3}^{+16.9}$} & \multirow{2}{*}{$16.1_{-0.4}^{+0.8}$} \\
   \rule[-0.12cm]{0cm}{4.8mm}(- - -)   &      -$i64.5_{-1.7}^{+2.8}$     &           &         &        \\ \hline
   \rule[-0.12cm]{0cm}{4.8mm}VII & $6872.7_{-8.6}^{+6.0}$  & \multirow{2}{*}{$ 1349.7_{-10.8}^{+29.1}$} & \multirow{2}{*}{$944.4_{-7.6}^{+19.9}$}  & \multirow{2}{*}{$14.5_{-0.3}^{+0.9}$} \\
   \rule[-0.12cm]{0cm}{4.8mm} (- + -)   &   -$i46.5_{-1.0}^{+2.1}$    &       &        &          \\ \hline
   \end{tabular}
      \caption{\label{tab:poles;3} Pole locations for Fit.II. } 
}
   \end{table}
In both Fits, only one resonance is found\footnote{Very recently, another state, $X(6600)$ is found in di-$J/\psi$ invariant mass spectra by CMS \cite{Zhang:2022toq}. In contrast, it is not clear in the 4$\mu$ ($J/\psi J/\psi$ and $J/\psi \psi(2S)$) spectra as measured by ATLAS \cite{Xu:2022rnl}. Here the resonant structure around 6600 MeV of LHCb is not obvious, and in practice, we do not find such a resonance, see Fig.\ref{Fig:fit}. }. For the coupled-channels case, Fit.~I, two poles are found in Riemann sheet (RS)-II and RS-III, with the quantum number $0^{++}$. The pole in RS-III is the one closest to the physical sheet, and the location is given as 
$M=6884.2^{+8.2}_{-4.0}$~MeV and  $\Gamma=58.2^{+3.0}_{-0.2}$~MeV. 
Because of the pole counting rule \cite{Morgan:1992ge,Dai:2011bs}, a pair of poles in RS-II and RS-III suggest that the $X(6900)$ should be a Breit-Wigner type particle. Meanwhile, this resonance contains at least four quarks, $cc\bar{c}\bar{c}$, and hence it is likely to be a compact tetraquark.
Its couplings to the $J/\psi J/\psi$,  $J/\psi \psi(2S)$ channels are given in Table~\ref{tab:poles;2}. The magnitudes of $g_1$ and $g_2$ are large and in the same order. It implies that both channels, $J/\psi J/\psi$ and $J/\psi \psi(2S)$ couple strongly to the $X(6900)$. 

Similarly, in the triple-channels case, we find four poles in RS-II, RS-III, RS-IV, and RS-VII, with the quantum number $0^{++}$. See Table~\ref{tab:poles;2}. The magnitudes of the residues,  $g_3$, are much smaller than that of $g_1$ and $g_2$. This confirms that the $J/\psi \psi(3770)$ channel contributes only a bit to the $X(6900)$ and can be ignored somehow.
The pole closest to the physical sheet locates in RS-IV, and it gives  $M=6861.0_{-8.8}^{+6.3}$~MeV and $\Gamma=129.0_{-3.4}^{+5.6}$~MeV, while the other three accompanying shadow poles are in RS-II, RS-III, and RS-VII, and they are not far away. According to the pole counting rule of triple channels~\cite{Morgan:1992ge,Dai:2012kf}, it again should be a Breit-Wigner particle, that is, a compact tetraquark.

\begin{figure}[htbp]
   \centering
   \includegraphics[width=0.98\linewidth, height=0.25\textheight]{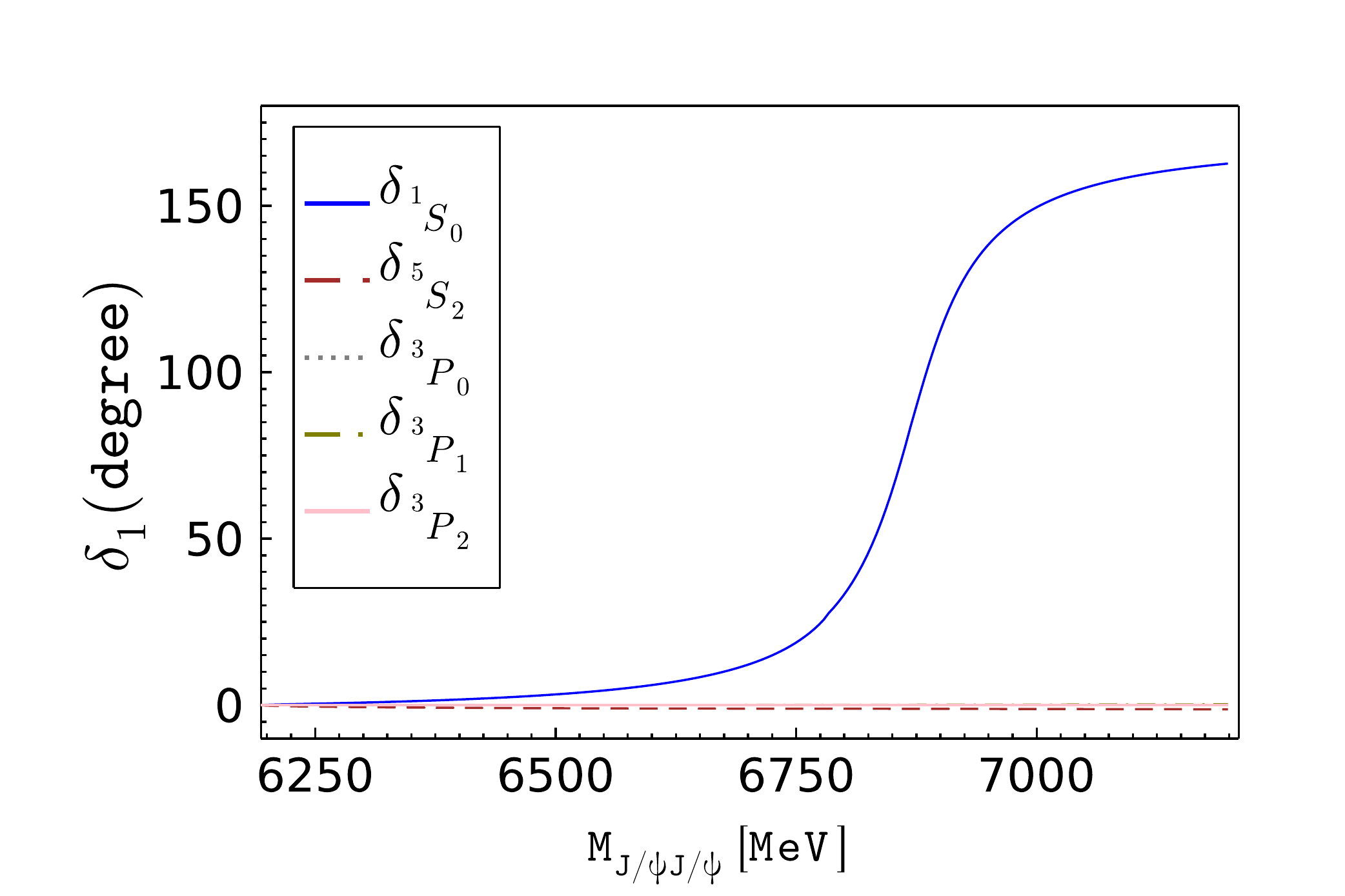}
   \caption{Phase shifts of each partial wave of Fit.~II. The one of Fit.~I is similar, and we do not show it. \label{Fig:ph} }
   \end{figure}

We produce the phase shifts of $\delta_1$ ($J/\psi J/\psi$) of each partial wave; See Fig.\ref{Fig:ph}.
As can be seen, the phase shift of the $^1S_0$ wave is smooth and looks very likely to be produced by a normal Breit-Wigner resonance. Others are small. Specifically, that of the $^5S_2$ wave is negative and quite flat, and that of the $^3P_1$ wave is positive and also contributes a small background, while the phase shifts of the other two partial waves are even more minor. Correspondingly, we do not find any poles in these partial waves. 
It again supports the $X(6900)$ to be a compact tetraquark.

\sectitle{Summary}
\label{Sec:IV}
In this letter, coupled-channel scatterings of $J/\psi J/\psi$-$J/\psi \psi(2S)$ and $J/\psi J/\psi$-$J/\psi \psi(2S)$-$J/\psi \psi(3770)$ are studied. The lowest-order effective Lagrangians are constructed, and the scattering amplitudes are calculated up to NLO. Partial wave decomposition is performed, and Pad\'e approximation is applied to restore unitarity.  
By fitting to the $J/\psi J/\psi$ invariant mass spectra measured by LHCb, we fix the couplings and extract out pole information of the $X(6900)$: $M=6884.2^{+8.2}_{-4.0}$~MeV and $\Gamma=58.2^{+3.0}_{-0.2}$~MeV for the coupled-channel case and $M=6861.0_{-8.8}^{+6.3}$~MeV and $\Gamma=129.0_{-3.4}^{+5.6}$~MeV for the triple-channel case. Its quantum number is likely to be $0^{++}$.
By pole counting rule and analysis on phase shifts of $J/\psi J/\psi$ scattering amplitudes, it is realized that the $X(6900)$ should be a Breit-Wigner-type particle (a compact tetraquark). It would be rather helpful if future experiments could measure the relevant angular distributions to refine this analysis.

\sectitle{Acknowledgements}
\label{Sec:V}
We thank Professors M. Shi and W. Shan  for helpful  discussions. This work is supported by Joint Large Scale Scientific Facility Funds of the National Natural Science Foundation of China (NSFC) and Chinese Academy of Sciences (CAS) under Contract No.U1932110, NSFC Grants with No.~11805059, No.~11675051 and No.~12061141006.

\bibliography{ref}

\end{document}